\documentclass[multphys,vecphys]{svmult}

\usepackage{makeidx}         
\usepackage{graphicx}        
\usepackage{multicol}        
\usepackage[bottom]{footmisc}

\makeindex             

\begin{document}

\title*{Multiple stars: designation, catalogues, statistics}
\author{A.~Tokovinin\inst{1} }

\institute{Cerro Tololo Inter-American Observatory, Casilla~603, La Serena, 
Chile
\texttt{atokovinin@ctio.noao.edu}}
%
\maketitle

\begin{abstract}
Discussion of  the designation of  multiple-star components leads  to a
conclusion that,  apart from components, we need  to designate systems
and centers-of-mass.  The  hierarchy is coded then by  simple links to
parent. This  system is  adopted in the  multiple star  catalogue, now
available  on-line.  A  short  review of  multiple-star statistics  is
given: the frequency of different multiplicities in the field, periods
of  spectroscopic sub-systems,  relative orbit  orientation, empirical
stability  criterion,  and  period-period  diagram with  its  possible
connection to formation of multiple stars.
\end{abstract}

\section{Designation of multiple stars}
\label{sec:1}

Actual  designations  of binary  and  multiple  stars are  historical,
non-systematic and confusing. Future space missions like GAIA, ongoing
searches for  planets, and other large projects  will greatly increase
the number of known multiple stars and components, making it urgent to
develop  a coherent and  unambiguous designation  scheme.  Recognizing
this need, IAU  formed a special group that held  its meetings in 2000
and 2003.  As  a result, IAU adopted the  designation system developed
for  the  Washington  Multiple  Star Catalogue  (WMC)  \cite{WMC},  an
extension of  the current designations  in the Washington  Double Star
Catalogue (WDS) \cite{WDS}.  Old  WDS designations will not change (to
avoid  further confusion),  new  component names  will be  constructed
hierarchically as  sequences of letters  and numbers, e.g.   Ab2.  The
designations reflect  hierarchical structure of  multiple systems, but
only to a  certain extent.  Future discoveries of  components at outer
or  intermediate  levels of  hierarchy  and  the  constraint of  fixed
designations  will inevitably lead  to situations  where names  do not
reflect the true hierarchy, which has then to be coded separately.

\begin{figure}[ht]
\centering
\includegraphics[width=10cm]{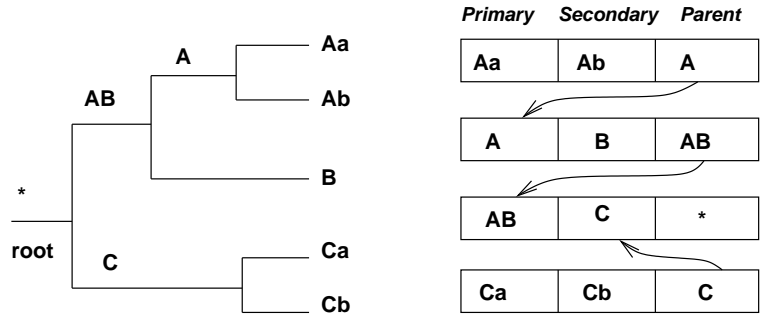}
\caption{Hierarchical structure  of a quintuple system  and its coding
  by  reference  to  parent.  Given the  components  designations  and
  references to  parent (right), the  hierarchical tree (left)  can be
  constructed automatically. }
\label{fig:1}    
\end{figure}

In the  future, many components  of multiple systems will  be resolved
into  sub-systems, so  that  A will  become  Aa and  Ab, for  example.
Nevertheless, current designations of these components (or rather {\em
super-components}, as they contain more than one star) will not become
obsolete, because past and future  measurements of these entities as a
whole will still  refer to the super-component names  such as ``A'' in
our  example. It  turns out  that the  concept of  super-components is
crucial to the  whole problem.  Once each super-component  has its own
designation,  unique   within  each  multiple  system,   we  can  code
hierarchies by simple reference to parent (Fig.~\ref{fig:1}).

The  WMC   designation  system  extended  to   normal  components  and
super-components   is  thus  logical   and  flexible,   permitting  to
accommodate new discoveries. Its  application is not free of ambiguity,
however, so  that a  common center (or  clearing house) will  still be
required. The WMC itself will hopefully play this role.

\section{Multiple-star catalogues}

Although multiple stars  are quite common, this fact  is not reflected
in the  catalogues.  A  researcher wishing to  study large  samples of
multiple stars has to compile his own lists or use the lists published
by others,  e.g. \cite{Fekel,DM}.  It is equally  possible to extract
multiple stars from binary  catalogues.  The current 9-th catalogue of
spectroscopic   binary  orbits,  SB9\footnote{http://sb9.astro.ulb.be}
\cite{SB9},   contains   multiplicity   notes   and   visual-component
designations.  Many visual multiples are listed in the WDS \cite{WDS},
but  the  physical relation  between  their  components  has not  been
studied systematically,  many are  simple line-of-sight projections
(optical).

The Multiple Star  Catalogue (MSC) \cite{MSC} is an  attempt to create
and  maintain  a  list  of  {\em  physical} systems  with  3  or  more
components.   The  MSC  is  essentially  complete  for  ``historical''
multiple  systems known  before  1996, with  the  exception of  visual
multiples  (only a fraction of  WDS  multiples have  been checked  for
physical relation).  The  completeness of the MSC (1024  systems as of
July  2005) is  less  evident  with respect  to  new discoveries.   An
on-line     interface     to      the     MSC     became     available
recently\footnote{http:/www.ctio.noao.edu/\~{}atokovin/stars/}.     The
MSC  contains estimates  of component  masses and  orbital parameters,
essential for statistical  studies, e.g.  \cite{Valtonen98}.  Several
observing programs have used the MSC to create their samples.  The new
designation system  is now implemented in the  MSC, hierarchical trees
like that in Fig.~1 are displayed on-line.

\section{Statistics of multiple stars}

\begin{figure}
\centering
\includegraphics[width=8cm]{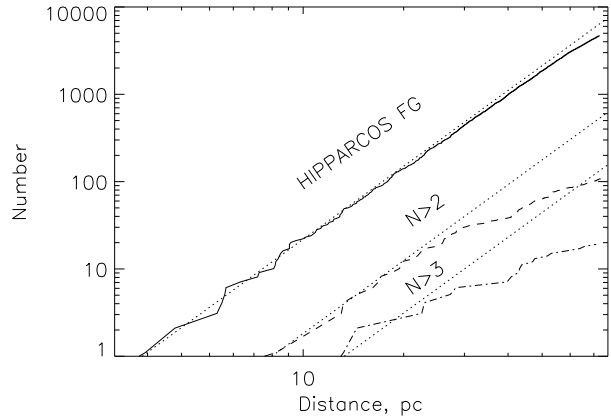}
\caption{ Completeness of the  multiplicity knowledge: number of stars
and systems of spectral types F and G ($0.5 < B-V < 0.8$) within given
distance.   Full  line:  objects  in the  HIPPARCOS  catalogue  follow
reasonably well the expected cubic  law (dotted line). Dashed line: F-
and G-type dwarfs  with 3 or more companions from  MSC also follow the
cubic law,  but only up  to a distance  of 30~pc (1/12th of  all stars
have 3 or  more components).  At 50 pc,  the estimated completeness is
only 40\%.  Dash-dot: quadruples and higher multiplicities, still very
incomplete.}
\label{fig:2}    
\end{figure}

\textbf{  Frequency of  multiple stars.}   The values  of multiplicity
fraction  given  in the  literature  are  often  confusing because  of
different definitions of this parameter.  Let $n_k$ be the fraction of
systems  with exactly  $k$ components,  and $a_k$  -- the  fraction of
systems  containing at  least  $k$ components  (i.e.  counting  higher
multiplicities  as well). Evidently,  $n_k =  a_k -  a_{k+1}$.  Batten
\cite{Batten} defines  the multiplicity ratio $f_k  = a_k/a_{k-1}$ and
argues that  $f_k \sim 0.25$  for $k \ge  3$.  This estimate  has been
confirmed with a larger sample  from the MSC \cite{T2001}.  It follows
immediately that $n_k = a_k(1 - f_{k+1})$.

Duquennoy  \&  Mayor \cite{DM91}  (DM91)  count  all binary  pairings,
irrespectively  of their  hierarchy. A  {\em companion  star fraction}
$CSF =  n_2 + 2n_3  + 3n_4 +  ... = 0.62$  can be inferred  from their
Fig.~7.   From Fig.~\ref{fig:2},  we  estimate $a_3  \approx 1/12$  (a
higher  estimate $a_3 =  0.2-0.25$ has  been given  in \cite{Merida}).
Assuming $f_k  = 0.25$,  we calculate $n_3$,  $n_4$, $n_5$,  etc., and
evaluate the contribution to the $CSF$ from pairings in multiple stars
as $2n_3  + 3n_4 +  4n_5 +...  =  0.26$.  Hence, the fraction  of pure
binaries in  the DM91 sample should  be $n_2 = 0.62-0.26  = 0.36$, the
fraction of  systems that are at  least binary is  $a_2 = n_2 +  a_3 =
0.44$, and the fraction of higher hierarchies with respect to binaries
is $f_3 = a_3/a_2 =0.19$.  A  smaller number $f_3 = 0.11 \pm 0.04$ (or
$a_3 = 0.05$) can be derived directly from the DM91 data \cite{T2001}.

\textbf{  Short-period sub-systems  and Kozai  cycles.}  Spectroscopic
binaries in the field seem to have a period distribution that smoothly
rises toward longer  periods in the range from  1 to 1000~d, according
to  several   independent  studies  \cite{DM91}.    In  contrast,  the
distribution of periods of spectroscopic sub-systems in multiple stars
shows a maximum at periods below 7~d \cite{TS02}.  This ``feature'' is
too sharp  to be  explained by selection  effects, and  the transition
period  is  suspiciously  similar   to  the  cutoff  period  of  tidal
circularization.   Dissipative  Kozai   cycles  are  the  most  likely
mechanism  that shortens  periods of  many (but  not  all) sub-systems
below 7~d.

\textbf{ Relative  orientation of orbits} has been  studied by Sterzik
\& Tokovinin \cite{SteTok}.  Only in  22 cases the {\em visual} orbits
of both  outer and inner  sub-systems are known. This  extremely small
sample has small ratios  of outer-to-inner periods $P_{\rm out}/P_{\rm
in}$  because at  long $P_{\rm  out}$  the time  coverage of  existing
visual  data  (about  200  yr)  is  still too  short  (in  fact,  many
long-period orbits  are uncertain or  wrong), while orbits  with short
$P_{\rm in}$ are difficult to  get for the lack of spatial resolution.
The  true  ascending  nodes  of  visual  orbits  are,  generally,  not
identified, further  complicating data interpretation.   The advent of
adaptive optics  and long-baseline interferometry  holds great promise
in extending this sample  significantly, mostly by resolving the inner
(spectroscopic)  sub-systems  in  visual  binaries  with  known  outer
orbits, e.g. \cite{Muterspraugh}.

Despite current  observational limitations,  it is already  clear that
the inner and outer orbits are neither coplanar nor completely random.
The directions  of their orbital angular momenta  are weakly correlated.
Such  correlation  could be  explained  by  dynamical  decay of  small
stellar groups  \cite{SteTok}.  However, alternative  explanations are
possible, too.  It  will be extremely important to  extend the studies
of relative orbit  orientation to larger samples and  to start probing
the orientations in different sub-groups.

\begin{figure}[ht]
\centerline{
\includegraphics[width=7cm]{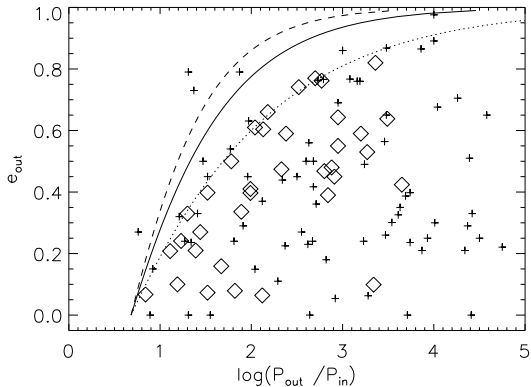}
\hspace{0.5cm}
\parbox[b]{4cm}{
\caption{Comparison  of  dynamical  stability  criteria  with  orbital
  parameters  of the  real systems:  eccentricity of  the  outer orbit
  $e_{\rm   out}$  (vertical   axis)  versus   period   ratio  $P_{\rm
  out}/P_{\rm  in}$  (horizontal axis).   The  full  line depicts  the
  dynamical  stability  criterion of  MA02,  the  dashed  line is  its
  modification  proposed  in \cite{SteTok},  the  dotted  line is  the
  empirical criterion.
\label{fig:3}    }
} }
\end{figure}


\textbf{ Empirical stability  criterion.}  Multiple systems where both
inner and  outer orbits are  known offer rich possibilities  for joint
analysis  of their  orbital parameters,  e.g.  checking  the dynamical
stability.   Theoretical and numerical  formulations of  the stability
criterion in the three-body problem  have been offered by many authors
and lead  to similar results.  I  take the latest work  of Mardling \&
Aarseth  \cite{MA2002}  (MA02)  as  representative and  compare  their
stability criterion with 120 real systems from the MSC (Fig.~3).  Some
systems  are,  apparently,  unstable.   However,  I  can  ignore  both
unreliable  outer orbits with  periods over  300~yr and  inner periods
shorter than 10d (likely affected by Kozai cycles), plotted as crosses
in  Fig.~3.   The remaining  systems  (diamonds)  nicely  fall in  the
stability zone.

When outer orbits are nearly  circular, the match between the data and
the MA02 criterion is impressive: all systems indeed have $P_{\rm out}/P_{\rm
in}  >   4.7$.   However,  eccentric  outer  orbits   deviate  from  the
theoretical  criterion  in  a  systematic  way.   The  {\em  empirical
stability criterion}  \cite{Merida} can  be described by  the relation
$P_{\rm out}(1-e_{\rm out})^{3}/P_{\rm in}>5$, whereas all theoretical
criteria  lead  to a  similar  relation  with $(1-e_{\rm  out})^{2/3}$
instead of cube. The reason of this discrepancy remains a mystery.

\begin{figure}[ht]
\centerline{
\includegraphics[width=7cm]{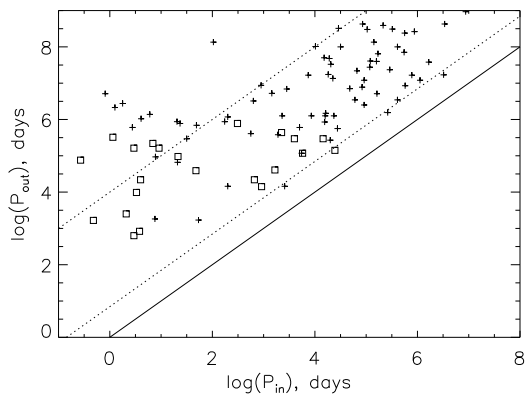}
\hspace{0.5cm}
\parbox[b]{4cm}{
\caption{ Periods of inner (horizontal axis) and outer (vertical axis)
sub-systems  in  a  sample  of  nearby late-type  multiples  from  the
MSC. Systems where  both periods are known from  orbital solutions are
plotted as  squares, in the  remaining systems (crosses) at  least one
period is estimated from the  separation. The full line corresponds to
equal periods, the dashed lines depict period ratios of 5 and 10\,000.
\label{fig:4}    }
} }
\end{figure}


\textbf{  Period-period  diagram.}  What  is  a  typical period  ratio
$P_{\rm  out}/P_{\rm  in}$  at  adjacent hierarchical  levels?   Fekel
\cite{Fekel}  found  it to  be  large, around  2000.  In  the MSC,  we
encounter all possible ratios allowed by the dynamical stability, i.e.
greater than 5.  However,  systems at intermediate hierarchical levels
often remain undiscovered, leading to wrongly estimated period ratios.
By restricting the sample to nearby (within 50~pc) late-type stars, we
reduce these errors  and begin to see the  true distribution of period
ratios (Fig.~4).

The  $P_{\rm out}  -  P_{\rm  in}$ diagram  reveals  some features  of
multiple-star  formation and  evolution. Interestingly,  period ratios
larger than 10\,000 are found only when $P_{\rm in} < 30$d, i.e. where
inner periods were likely shortened by some dissipative mechanism like
Kozai cycles with  tides.  The only exception to  this rule (the point
at the top) is Capella, a pair  of giants on a 100-d circular orbit in
a quadruple  system. It  is very likely  that all multiple  stars have
been  formed   with  the  period  ratio  $P_{\rm   out}/P_{\rm  in}  <
10\,000$. Some  inner periods  were then shortened  by tidal  or other
dissipative  processes.  In  this  perspective, Capella  had a  rather
eccentric initial orbit  with a period of few tens  of years which has
been shortened and circularized when its components became giants.

\section{Conclusions}

Cataloguing of multiple systems,  however boring it might seem, offers
interesting  insights  into formation  and  evolution  of stars.   New
powerful   observing  techniques  (adaptive   optics,  interferometry,
precise  radial velocities)  should now  be applied  to  large stellar
samples in order  to ``fix'' the multiplicity statistics  in the solar
neighborhood and beyond.



\printindex
\end{document}